\documentclass[10pt,conference,compsocconf]{IEEEtran}
\ifCLASSOPTIONcompsoc
  \usepackage[nocompress]{cite}
\else
  \usepackage{cite}
\fi
\ifCLASSINFOpdf
\else
\fi
\usepackage{url}

\PassOptionsToPackage{hyphens}{url}
\usepackage{makecell}
\usepackage{multirow}
\usepackage{amsmath}
\usepackage{mathtools}
\usepackage{graphicx}
\graphicspath{ {images/} }
\usepackage[font=normalsize]{subfig}
\usepackage{algorithm}
\usepackage{algpseudocode}

\algnewcommand{\LineComment}[1]{\State \(\triangleright\) #1}

\begin{document}
%
\title{\huge Parallel Detection for Efficient Video Analytics at the Edge}

\author{\IEEEauthorblockN{Yanzhao Wu, Ling Liu}
\IEEEauthorblockA{School of Computer Science\\
Georgia Institute of Technology\\
Atlanta, Georgia
}
\and
\IEEEauthorblockN{Ramana Kompella}
\IEEEauthorblockA{Cisco Systems, Inc.\\}
{170 West Tasman Dr.} \\
San Jose, California
}


%


\maketitle

\begin{abstract}
Deep Neural Network (DNN) trained object detectors are widely deployed in many mission-critical systems for real time video analytics at the edge, such as autonomous driving, video surveillance, Internet of smart cameras. A common performance requirement in these mission-critical edge services is the near real-time latency of online object detection on edge devices. However, even with well-trained DNN object detectors, the online detection quality at edge may deteriorate for a number of reasons, such as limited capacity to run DNN object detection models on heterogeneous edge devices, and detection quality degradation due to random frame dropping when the detection processing rate is significantly slower than the incoming video frame rate. This paper addresses these problems by exploiting multi-model multi-device detection parallelism for fast object detection in edge systems with heterogeneous edge devices. First, we analyze the performance bottleneck of running a well-trained DNN model at edge for real time online object detection. We use the offline detection as a reference model, and examine the root cause by analyzing the mismatch among incoming video streaming rate, the video processing rate for object detection, and the output rate for real time detection visualization of video streaming. Second, we study performance optimizations by exploiting multi-model detection parallelism. We show that the model-parallel detection approach can effectively speed up the FPS detection processing rate, minimizing the FPS disparity with the incoming video frame rate on heterogeneous edge devices. We evaluate the proposed approach using SSD300 and YOLOv3 (pre-trained DNN models) on benchmark videos of different video stream rates. The results show that exploiting multi-model detection parallelism can speed up the online object detection processing rate and deliver near real-time object detection performance for efficient video analytics at edge.
\end{abstract}


%
\IEEEpeerreviewmaketitle

\section{Introduction}
As the high speed Internet continues to advance and evolve, our physical, social and cyber worlds are interweaving into a gigantic network of connected people, connected devices, and connected things. Soon we will be walking on smart streets, commuting with smart transportation, living in smart home, working in smart office buildings and smart manufacturing. This technological evolution trend will not only result in unprecedented growth of multi-modal digital data, but also drive network bandwidth and network resources to their limits, e.g., managing the payloads of transporting video data captured from a massive network of surveillance cameras or drones in real time 24 by 7. Edge computing has emerged as a popular distributed computation paradigm, which represents a new wave for computation offloading to where data is collected, and a new distributed system architecture for processing, learning and reasoning from dispersed data. 

Many edge systems today have employed DNN-trained object detection models as a critical component for mission-critical video analytic tasks at edge, such as autonomous driving~\cite{autonomous-driving-object-detection}, object tracking~\cite{object-tracking-real-time}, intrusion detection and surveillance monitoring for networks of cameras~\cite{video-surveillance-object-detection} in smart buildings, smart streets and smart cities. 

\noindent {\bf Object detection for video analytics.} A video stream can be decomposed into a sequence of video frames upon capturing. Offline frame processing and online frame processing represent two spectrums of object detection for video analytics. SSD~\cite{ssd} and YOLOv3~\cite{yolov3} are representative DNN-powered object detection algorithms popularly used for video analytics at edge. Object detectors pre-trained using DNN algorithms such as SSD~\cite{ssd} and YOLOv3~\cite{yolov3} can detect and localize objects by performing detection processing to identify objects with their bounding boxes and class labels in an input video frame, and then output these detection processed frames via visual display at edge in the original sequence of the incoming video frames. 
Offline detection and online detection represent two ends of the spectrum for video frame processing workflows in the context of incoming video frame rate, the detection processing rate and the output display rate. 

\begin{figure}[h!]
\centering
    \subfloat[Offline Processing]{
    \centering
    \includegraphics[width=0.45\textwidth]{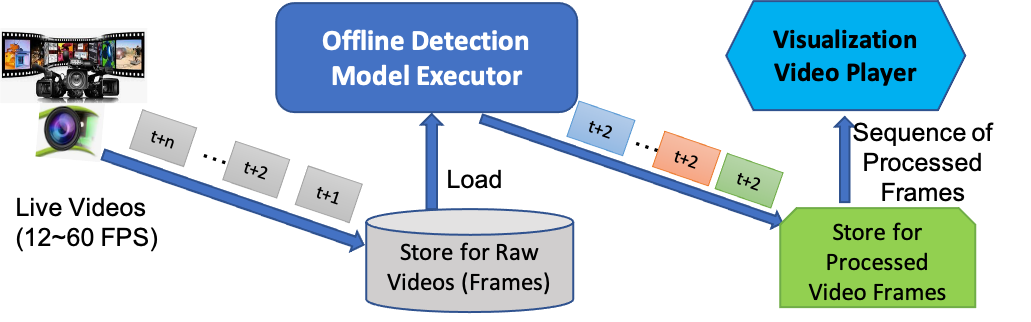}
    \label{fig:offline-workflow}
    }\newline
    \subfloat[Online Processing]{
    \centering
    \includegraphics[width=0.45\textwidth]{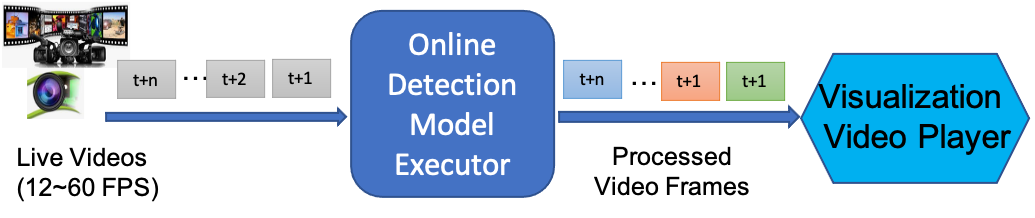}
    \label{fig:online-workflow}
    }
    \caption{Offline vs. Online End to End Object Detection Workflows}
    \label{fig:offline-online-workflow}
\end{figure}

The {\it offline processing} will perform the object detection on the incoming video frames one at a time regardless the incoming video rate in in the number of frames per second (FPS). It is used as a detection processing reference model for zero-frame drop. Once all the incoming video frames are processed for object detection using a pre-trained object detector such as SSD~\cite{ssd} and YOLOv3~\cite{yolov3}, the output of the processed frames will be sorted by the temporal sequence of the incoming video frames prior to visualization and presentation via a video player on an edge device. Figure~\ref{fig:offline-workflow} provides an visual illustration. First, the live video is captured and stored in a persistent storage, such as a solid-state drive. Then, each stored video is uploaded to the offline detection model executor, which performs object detection one frame at a time regardless the real time video FPS rate from the video camera or other video capturing equipment. Next, upon completing the processing of all frames of the input video, the processed video frames are sorted in the original temporal sequence of the input video to produce the detection output video, which will feed into the edge device for visualization and presentation via a video player. The offline processing time can be measured by the multiplication of the per frame processing time with the total number of frames. Alternatively, one can also output the detection processed frames as they become available and follow the temporal sequence of input frames. The main advantage of this offline detection is its guarantee of zero frame dropping during detection processing and the end to end video analytics workflow. However, we argue that it cannot be adopted in real time video analytics on edge devices when the real time detection processing rate is much slower than the incoming video streaming FPS rate. 

In contrast, an {\it online object detection} edge system is by design a detection streaming system, which will continuously feed the incoming video frames one after another to the online object detection model executor, regardless whether the detection processing is able to finish the current per-frame processing. This may result in random frame dropping. Although it will output the detection processed frames as they become available for visualization and presentation via a video player, instead of waiting for all frames to be processed and sorted. Figure~\ref{fig:online-workflow} sketches an end-to-end online detection workflow. In comparison, online detection has a number of unique characteristics. First, when incoming video streaming rate is much faster than the detection processing FPS rate, the edge detection may experience malfunction, such as random dropping of large frames. Second, even though online detection will output the detection processed frames as they become available, the slow detection processing rate may lead to low output FPS rate, rendering the edge video analytics failing to deliver near real time performance. 

\noindent {\bf Challenges for online object detection.\/} Given a live video stream rate, say 30 frames per second (FPS), and heterogeneous edge devices with different model execution capacities, it is challenging for online object detection model executor to deliver near real-time FPS processing performance (e.g., the detection processing rate of 30 FPS) for a number of reasons.  {\it First,} edge devices tend to be heterogeneous in computation, communication, storage, and energy consumption. Hence, running DNN object detection models on edge devices may result in a large variation in terms of detection processing rates in FPS. Performance degradation due to detection processing delay includes random frame dropping when the incoming video stream is faster than the detection processing rate. When the gap between the incoming FPS and the detection processing FPS becomes significantly large, the edge system will experience slow output stream with glitches and lagging issues and poor detection quality, such as low mAP (mean average precision) for object detection quality over the total frames of the input video.
{\it Second,} DNN trained object detection models are computationally intensive for many edge devices~\cite{faster-rcnn,ssd,yolov3,fast-yolo}. Although recent studies have shown Faster RCNN~\cite{faster-rcnn}, SSD~\cite{ssd} and YOLOv3~\cite{yolov3} can deliver real-time or near real-time detection performance on medium or high-end GPUs, few studies have dedicated on delivering real-time or near real-time performance for online object detection on edge devices equipped with edge AI hardware, such as the Intel Neural Computing Stick 2 (NCS2)~\cite{intel-ncs2}, Raspberry Pi~\cite{raspberry-pi}, NVIDIA Jetson~\cite{nvidia-jetson} and Google Coral~\cite{google-coral}. These edge AI hardwares tend to employ custom software frameworks, such as Intel OpenVINO~\cite{openvino}, NVIDIA EGX~\cite{nvidia-egx} and Google TensorFlow Lite~\cite{tensorflow-lite} to run compressed DNN object detection models on edge devices for supporting Edge AI applications. Although powered by such edge AI hardware, edge devices with different computation capacity can run pre-trained object detection models for edge video analytics, we will show in this paper that the detection processing performance in FPS on such edge AI hardware is slower compared to that of medium or high-end GPUs, a root cause for the performance problems due to the mis-match between the fast incoming video stream rate and the slow detection processing rate.

\noindent {\bf Scope and contributions.} We argue that an important quality of service requirement for edge systems is to provide near real-time performance guarantee for object detection on edge devices of heterogeneous resources. In this paper, we exploit edge detection parallelism for fast object detection in edge systems with heterogeneous edge devices, aiming to significantly enhance the detection processing rate while minimizing random dropping, enabling high detection quality in the presence of limited capacity to run DNN object detection models on heterogeneous edge devices. First, we analyze the performance bottleneck of running a well-trained DNN model at edge with Intel NCS2~\cite{intel-ncs2} installed for real time online object detection. We use offline detection as a reference model for zero-frame dropping, and examine the root cause of random dropping when the detection processing rate is significantly slower than the incoming video streaming rate. Second, we address the detection processing throughput problem, we present a performance optimization technique by exploiting multi-model multi-device detection parallelism for fast object detection at edge. Our parallel model detection approach consists of a careful combination of a model parallel scheduler and a model parallel sequence synchronizer. It can effectively speed up the FPS detection processing rate, minimizing its FPS disparity with the incoming video frame rate. 
Experiments are conducted on two benchmark video datasets~\cite{MOTChallenge2015} with diverse incoming FPS rates using well-trained DNN object detection models (SSD~\cite{ssd} and YOLOv3~\cite{yolov3}). We show that by exploiting model detection parallelism and leveraging multiple AI hardware devices at edge, we can effectively speed up the online object detection processing rate and deliver near real-time object detection performance for efficient video analytics at edge.

\begin{figure*}[h!]
\centering
    \subfloat[Frame 64]{
    \centering
    \includegraphics[width=0.249\textwidth]{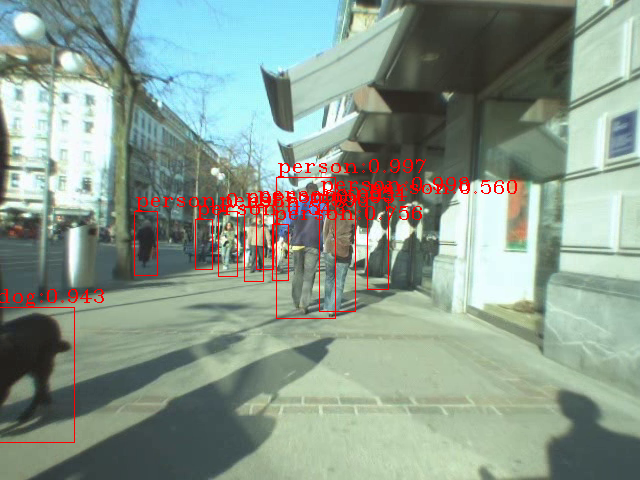}
    \label{fig:eth-sunnyday-video-frames-no-drop-64}
    }
    \subfloat[Frame 65]{
    \centering
    \includegraphics[width=0.249\textwidth]{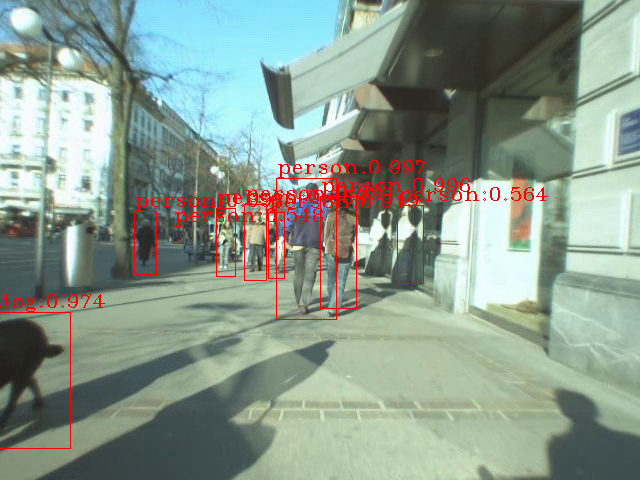}
    \label{fig:eth-sunnyday-video-frames-no-drop-65}
    }
    \subfloat[Frame 66]{
    \centering
    \includegraphics[width=0.249\textwidth]{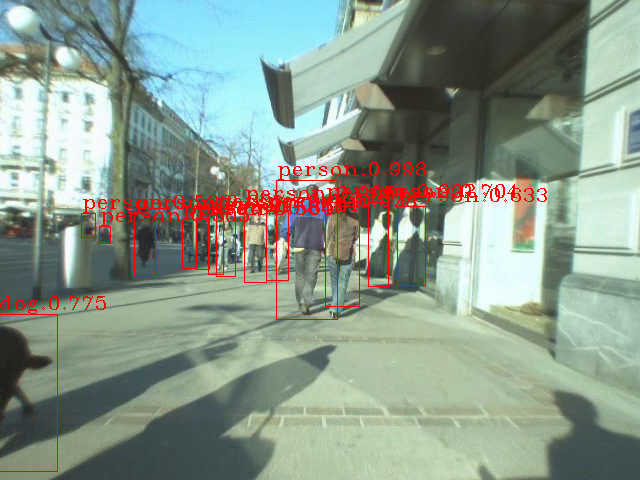}
    \label{fig:eth-sunnyday-video-frames-no-drop-66}
    }
    \subfloat[Frame 67]{
    \centering
    \includegraphics[width=0.249\textwidth]{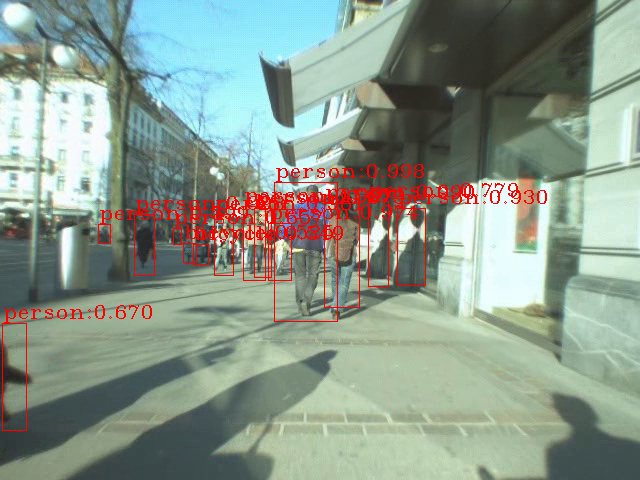}
    \label{fig:eth-sunnyday-video-frames-no-drop-67}
    }
    \caption{Frame 64$\sim$67 without dropping frames (ETH-Sunnyday, Single NCS2, YOLOv3, Processing FPS=2.5, mAP=86.9\%)}
    \label{fig:eth-sunnyday-video-frames-no-drop}
\end{figure*}

\begin{figure*}[h!]
\centering
    \subfloat[Frame 64]{
    \centering
    \includegraphics[width=0.249\textwidth]{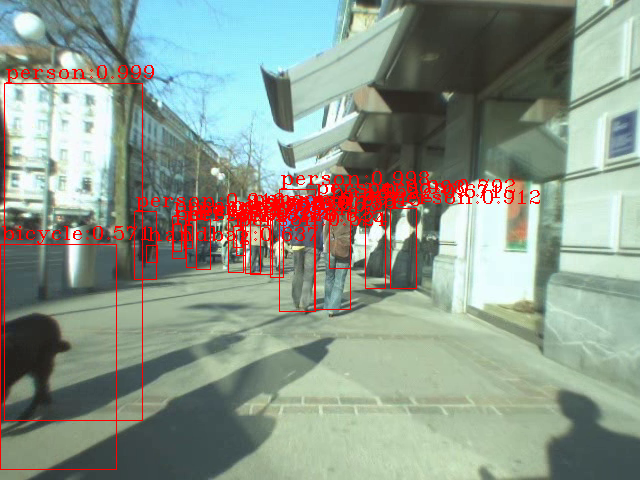}
    \label{fig:eth-sunnyday-video-frames-drop-64}
    }
    \subfloat[Frame 65]{
    \centering
    \includegraphics[width=0.249\textwidth]{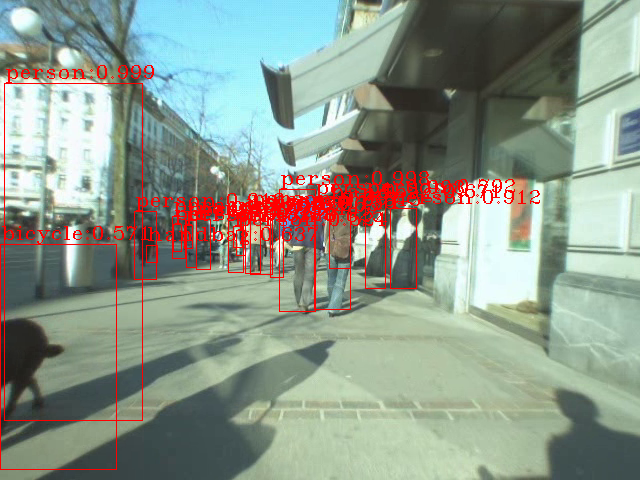}
    \label{fig:eth-sunnyday-video-frames-drop-65}
    }
    \subfloat[Frame 66]{
    \centering
    \includegraphics[width=0.249\textwidth]{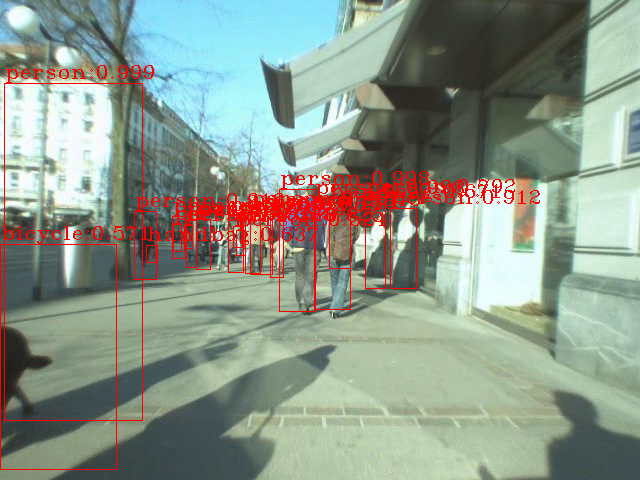}
    \label{fig:eth-sunnyday-video-frames-drop-66}
    }
    \subfloat[Frame 67]{
    \centering
    \includegraphics[width=0.249\textwidth]{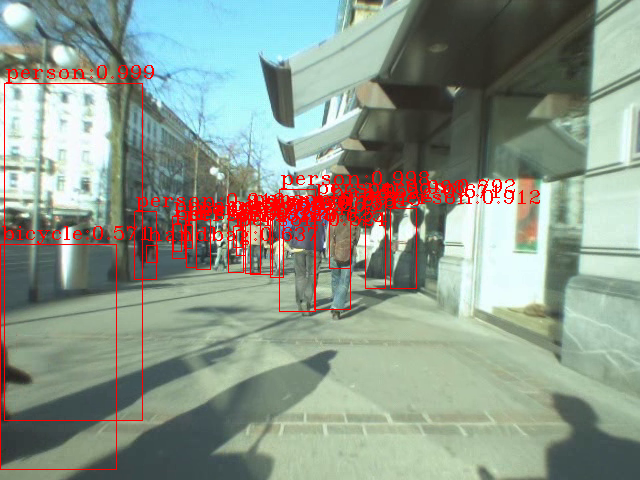}
    \label{fig:eth-sunnyday-video-frames-drop-67}
    }
    \caption{Frame 64$\sim$67 with dropping frames (ETH-Sunnyday, Single NCS2, YOLOv3, Processing FPS=14.0, mAP=66.1\%)}
    \label{fig:eth-sunnyday-video-frames-drop}
\end{figure*}

\section{Online Object Detection}
\subsection{Problem Statement}
We study the real time object detection at edge based on the following assumption. First, let the incoming video rate of a live video stream be $\lambda$ frames per second (FPS), 
and let the detection throughput rate of an object detector deployed on an edge device be $\mu$ frames per second. When $\mu \ge \lambda$, it indicates that the incoming video stream at $\lambda$ rate can be processed in real time with no frame dropping. However, in many cases, given the limited computing resources on an edge device, including the AI hardware for running the pre-trained DNN object detection model, the detection processing rate $\mu$ tends to be much smaller than the live video streaming rate $\lambda$. When $\mu << \lambda$, e.g., $\lambda=30$ FPS but $\mu=3$ FPS, we observe that even with well-trained object detectors (SSD or YOLOv3) deployed on an edge device with an AI hardware such as Intel NCS2 (Neural Compute Stick 2)~\cite{intel-ncs2}, the actual online object detection at edge may suffer from low frame processing rate, denoted by $\sigma$. An obvious performance optimization goal is to speed up $\sigma$ as much as possible by maximizing resource and computation capacities. Consider the two ends of the spectrum in implementing and optimizing the video frame processing rate $\sigma$ for object detection at edge: (1) if we want to maintain zero frame dropping, then we limit $\sigma$ to the offline detection processing baseline rate $\mu$, i.e., $\sigma = \mu$. When the incoming video stream rate $\lambda$ is much higher than $\mu$, i.e., $\mu << \lambda$, the online detection will suffer from the low video processing rate ($\sigma$) and hence low throughput in detection output FPS; and (2) if we want to speed up the actual video processing rate $\sigma$ at edge, we would need to develop resource and computation optimizations such that we can offer fast detection processing rate, that is either the same as the live video stream input rate $\lambda$, e.g., $\sigma = \lambda$, or increasing $\sigma$ towards $\lambda$.

{\bf Na\"ive Approach.\/} When the actual processing rate of the detection model on an edge device is much lower than the input video streaming rate $\lambda$, e.g., say $\mu=2.5$ FPS whereas $\lambda=30$ FPS, if the online detection scheduler simply sends the incoming video frames to the AI hardware for object detection as they arrive at $\lambda$ rate, then the online detection executor may end up dropping some video frames randomly at approximately $(\lambda - \mu)$ frames per second. This translates to dropping $(\lambda/\mu - 1)$ frames on average for every frame processed by the object detector running on the AI hardware attached to an edge device. This may result in decreasing the mean average precision (mAP) for object detection quality measured over all frames of the input video. Such performance degradation can be further aggravated when the mismatching gap between the input video stream rate $\lambda$ and the detection processing rate at edge is much larger due to  a larger number of randomly dropped frames (see experimental results in Section~\ref{section:experimental-analysis}). 


\subsection{Illustrative Examples and Visualization}
In this subsection, we use a benchmark video ETH-Sunnyday from the MOT-15 dataset~\cite{MOTChallenge2015}, which has a total of 354 frames, and is taken from a moving camera at 14 FPS with a resolution of 640$\times$480. We install YOLOv3 object detection model trained using 
DarkNet-53~\cite{yolov3} on an edge server with Intel NCS2~\cite{intel-ncs2}. We first resize the input video frame to the input size of the object detection model, that is 416$\times$416$\times$3 for YOLOv3. Then, the online object detection module running YOLOv3 will process each resized video frame by performing perform object detection inference, followed by post-processing, such as non-maximum suppression (NMS)~\cite{faster-rcnn,ssd,yolov3}, for computing the bounding box, the class label and the object confidence for each detected object. By applying post-training quantization to use 16 bit half-prevision floating point format (FP16), YOLOv3 is converted into the FP16 data type in order to be deployed on Intel NCS2 sticks with a smaller model size of 119MB instead of over 200MB.

Figure~\ref{fig:eth-sunnyday-video-frames-no-drop} shows the object detection results for four consecutive frames when we use the frame processing rate that will incur zero frame dropping as done in the offline detection, i.e., keeping a large buffer that is able to hold the input video frames at $\lambda$ rate, while scheduling one frame at a time to ensure no frame dropping, resulting in very low detection processing rate $\sigma$ with $\sigma << \lambda$. Even though in this case, the object detector can successfully identify the objects in each frame, the actual video processing rate $\sigma$ is only 2.5 FPS, which is 5.6x slower than the original ETH-Sunnyday live video stream rate of 14 FPS ($\lambda=14$). We also provide a visual comparison of this online detection processing with no frame dropping (i.e., $\mu=2.5$ FPS) with the original input live video stream rate. The visualization shows the shape contrast in \url{https://github.com/git-disl/EVA}, our EVA project site.

Figure~\ref{fig:eth-sunnyday-video-frames-drop} shows the online object detection processing, which aims to approximate the input video stream rate, for the video ETH-Sunnyday using the same four consecutive frames as shown in Figure~\ref{fig:eth-sunnyday-video-frames-no-drop}. Here the live video stream rate of 14 FPS is used to feed video frames to the online object detection executor (recall Figure~\ref{fig:online-workflow}). In this case, the online object detection executor experiences random frame dropping at the rate of 5 frames on average for each detection processed frame, i.e., $\lceil 14/2.5 - 1 \rceil = 5$. The random frame dropping happens when the object detection scheduler is sending incoming frames to the object detection executor at the input video rate of 14 FPS ($\lambda=14$), while its actual detection processing throughput is only 2.5 FPS (i.e., $\sigma=2.5 << \lambda$). By comparing Figure~\ref{fig:eth-sunnyday-video-frames-drop} with random dropping of input frames and Figure~\ref{fig:eth-sunnyday-video-frames-no-drop} with zero frame dropping, it is visually clear that in the four consecutive frames (64 to 67) of Figure~\ref{fig:eth-sunnyday-video-frames-drop}, a few bounding boxes are misaligned, and some buildings are mislabeled as person or bicycle (e.g., those on the left). As a result, the mAP of this online detection executor is reduced from 86.9\% to 66.1\% (see more detailed experimental results in Section~\ref{section:experimental-analysis}). Readers can find a visualization of the comparison on our GitHub project page (\url{https://github.com/git-disl/EVA}), showing the negative effects of random frame dropping on both FPS and mAP of the detection outputs through visual comparison of input video stream with the detection output video stream.

\begin{figure*}[h!]
\centering
    \includegraphics[width=0.75\linewidth]{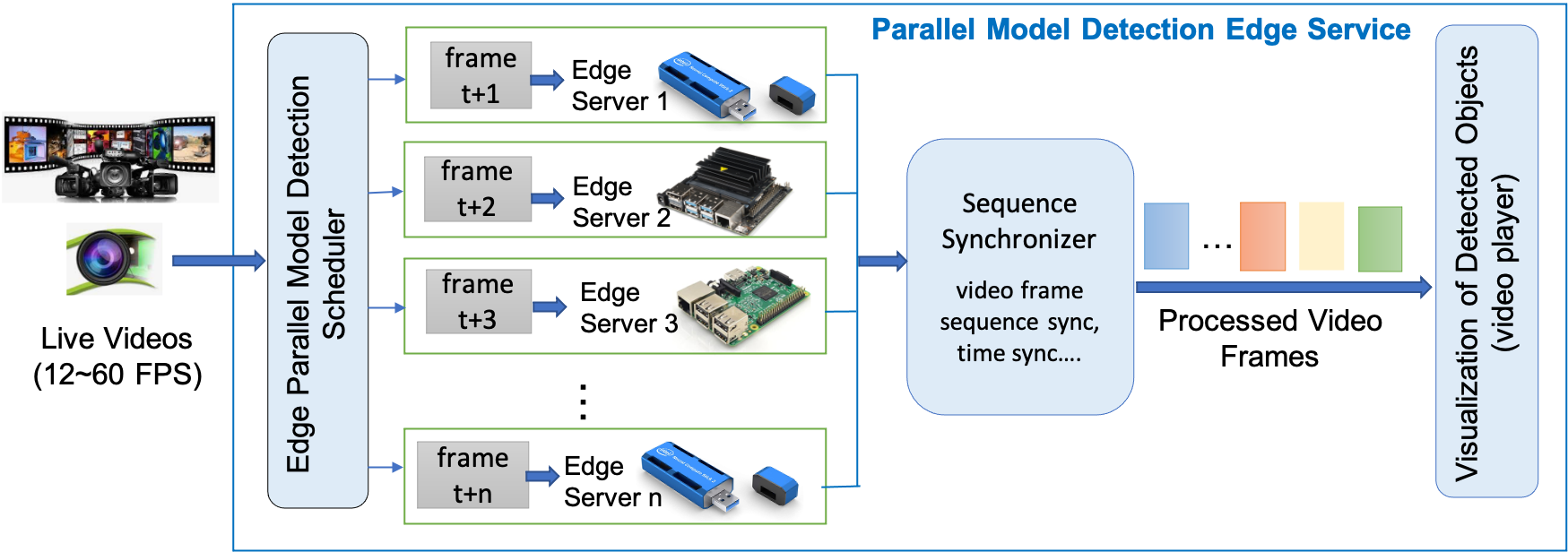}
    \caption{An Architectural Overview for Parallel Object Detection in EVA}
    \label{fig:model-parallel-workflow}
\end{figure*}

\section{Multi-Model Detection Parallelism}
With heterogeneous edge devices and different AI hardware processing capabilities, edge video analytics needs efficient mechanisms to cope with the mismatch between the input video streaming rate ($\lambda$) and the actual detection processing throughput rate ($\sigma$) on an edge device with a given AI hardware and pre-trained object detection model. In this paper, we address this mismatch problem by exploiting multi-model detection runtime parallelism. Put differently, we propose to design and implement an edge system for fast object detection by delivering the video processing rate $\sigma$ in near real-time. We aim to optimize the runtime detection processing efficiency by reducing the gap from the detection processing rate $\sigma$ to the given input live video streaming rate $\lambda$, while maintaining high accuracy performance in mean average precision (mAP) for efficient online object detection in edge systems.

\subsection{Solution Approach: An Overview}
Our multi-model parallel detection framework consists of three main functional components, which work in concert to explore the runtime detection parallelism at edge and to speed up the overall frame processing rate for object detection by reducing the performance gap with respect to the live input video stream rate $\sigma$. Figure~\ref{fig:model-parallel-workflow} shows a sketch of the overall workflow for fast online object detection by exploiting multi-model parallelism. The multi-model detection scheduler and the sequence synchronizer are the two core components, which work together to enable parallel detection processing of multiple input video frames at the same time while maintain the temporal order of the detection processed frames according to the original input video stream sequence. 

The multi-model parallel detection scheduler will first determine the number of parallel detection models to use for delivering the near real-time frame processing performance, denoted by $\sigma_{P}$, based on the input video stream rate $\lambda$ and the average single model detection processing rate $\mu$. Then the scheduler will create parallel execution thread pool, one for each model. Finally, the scheduler will choose an adequate scheduling algorithm based on the configuration of the edge system and its edge devices. There are several ways to exploit multi-model parallelism in an edge system, and one often complements another. First, we can explore multiple models running on one edge device by installing multiple AI hardware sticks as attachments, such as Intel NSC2 sticks~\cite{intel-ncs2}. Intel NSC2 is a hardware accelerator for deploying deep learning at the edge, which is built on the Intel Movidius VPU (Vision Processing Unit) and a dedicated neural compute engine with 4GB memory. It has the USB plus and play feature, enabling its wide deployment on many edge devices for accelerating DNN inference, supported by Intel OpenVINO software framework~\cite{openvino}. Note that 
the implementation of multi-model on one edge device will focus primarily on optimizing the model execution parallelism for maximizing the overall performance speed-up. Second and alternatively, we can explore multiple edge devices with one AI hardware like NCS2 attached to each device to run an object detection model. Additional performance factors need to be considered for this approach. For example, we need to design an algorithm to find sufficient edge devices within geographical proximity or with fast network connection to the leader edge device which initiates this video detection job. Then we need to define optimal workload balance for achieving overall performance speed-up when multiple nearby edge devices have initiated their own video detection jobs. The third alternative is to consider edge devices with heterogeneous AI hardware attachments, each is capable of running detection models of different size and complexity or different number of detection models. In summary, the first approach serves as a baseline, as it runs multiple detection models in parallel on one edge device attached with a USB 3.0 hub connecting multiple AI hardware sticks. The second alternative uses multiple nearby edge nodes to run multiple detection models instead, with each edge node attached with one AI hardware stick, running one object detection model. The third alternative addresses heterogeneous edge devices by implementing a hybrid of the first two approaches.

In the remaining of this section, we will describe our parallel multi-model detection architecture and algorithms in terms of the first design alternative, focusing on the design decision on the number of parallel detection models (the parameter $n$), the choice of parallel detection scheduling algorithms and the scheduler implementation, and the sequence synchronizer for achieving real time or near real time processing throughput $\sigma_{P}$. We report our experimental evaluation in Section~\ref{section:experimental-analysis}, with different edge node configurations and up to seven Intel NCS2 sticks as AI hardware attachments to an edge device.


\subsection{Determining the Parallel Detection Parameter $n$} 
\label{parameter_n}
The first task performed by our parallel object detection scheduler is to determine the number of detection models ($n$) to use in order to  achieve the desired online detection performance in terms of both frame processing rate FPS and mAP. The goal is to choose the right setting of the parallel detection parameter $n$ such that it is cost efficient. Too small $n$ may not be sufficient to achieve near real time detection processing throughput in FPS, whereas too large $n$ may result in resource waste when the parallel detection processing capacity $n\times \mu$ is far beyond the input video stream rate $\lambda$, i.e., $n\times \mu >> \lambda$. 

There are several factors impacting the choice of $n$, such as the computation capacity of the AI hardware attached to an edge device, the size and complexity of the pre-trained object detection model, the CPU and memory of the edge device, assuming that the number of AI hardware attachments via a USB hub connecting to an edge device is homogeneous. 
Let $\sigma_{P}$ denote the detection processing rate achieved by parallel execution of the $n$ detection models. 
One ideal scenario will be when the parallel object detection can scale linearly with the number of parallel models (or edge devices). As a result, the overall frame processing rate $\sigma_{P}=n\mu$, where $\mu$ is the per model detection processing rate, or the average detection processing rate of the $n$ detectors. $\sigma_{P}=\sum_{i} \mu_i$ when different models and/or heterogeneous edge devices are used, and each with its specific detection processing rate $\mu_{i}$ ($1\leq i\leq n$).
If we want to match or approach the original input video stream rate $\lambda$, we can set $n = \lceil \lambda / \mu \rceil$, which ensures that $\sigma_{P}=n \mu \geq \lambda$ holds. With this approach, we can  determine the right $n$ detection models to process the incoming video stream with zero frame dropping, because the parallel execution of $n$ detection models will be sufficient for online object detection in real time and no frame dropping will be in this ideal scenario. Given that the comfort of human perception for street view of walking pedestrians is around 10-30 FPS, one may further relax the ideal scenario and set $n$ in the range of [$\lceil 10/ \mu \rceil$, $\lceil \lambda / \mu \rceil $]. 
From observations made using our first edge system prototype, implemented in C++ on top of OpenVINO~\cite{openvino} and OpenCV~\cite{opencv}, we confirm that when the input video stream rate $\lambda$ is higher than 12 FPS, it is effective to choose $n$ from the range of [$\lceil 10/ \mu \rceil$, $\lceil \lambda / \mu \rceil$] instead of using the conservative setting of $n \geq \lceil \lambda / \mu \rceil$, if we aim at achieving near real-time object detection performance in FPS with high detection accuracy in mAP.

Recall the benchmark live video ETH-Sunnyday from the MOT-15-dataset~\cite{MOTChallenge2015} taken from a moving camera at 14 FPS (Figure~2 and Figure~3), if the detection processing rate for no-frame dropping ($\mu$)  is 2.5 FPS when running YOLOv3, and if we want to determine the parameter $n$, aiming to closely match the input video streaming rate $\lambda = 14$, we can achieve the parallel detection processing rate $\sigma_{P}$ by choosing $n$ from the range of ($\lceil 10/2.5 \rceil$,  $\lceil 14 / 2.5 \rceil$), which is [$4, 6$]. We get $\sigma_{P} = 10$ FPS when we choose $n=4$ by running four detection models in parallel, achieving the detection processing throughput close to the incoming video stream rate $\lambda$, which is 14 FPS, and mAP of 86.5\%. We call this choice of $n$ the near real time parallel detection approach. Alternatively, if we choose the upper bound of the range, which is $n=6$, then we achieve $\sigma_{P} = 14.8$ FPS with mAP of 86.9\%, indicating that by running 6 detection models in parallel, the detection processing throughput can match or slightly exceed the input video stream rate $\lambda=14$ with high mAP. We refer to this approach as a conservative real time parallel detection approach. Readers may refer to Table~\ref{table:exp-multi-ncs2-stick-eth-sunnyday} in Section~\ref{section:experimental-analysis} for more details.



\subsection{Parallel detection scheduling algorithms}
The second task performed by the parallel detection scheduler is also a part of the initialization process. It needs to choose a scheduling algorithm as a part of the runtime configuration of the scheduler. Concretely, upon receiving the input video stream at $\lambda$ rate, an edge device relies on its object detection task scheduler to launch $n$ threads to run $n$ object detection models in parallel. The assignment of incoming frames to $n$ detection models is done by a scheduling algorithm. There are three categories of parallel execution scheduling  algorithms: Round Robin (RR) or weighted round robin, First Come First Serve (FCFS), and performance aware proportional scheduler. 

A {\bf round robin scheduler} will assign incoming video frames to the $n$ detection models, one at a time, in a pre-defined round robin order at the input stream rate $\lambda$. For example, consider the input video stream $V$, the RR scheduler will dedicate a separate thread to receive the incoming video frames and place them into a shared parallel execution queue $Q$. At the same time, the RR schedule will launch $n$ dedicated threads, each runs one of the $n$ object detection models in parallel. Each thread will fetch one video frame from the shared queue $Q$ and send it to its corresponding object detection model for parallel detection processing. Upon completing one round of input frame to detection model assignment for all $n$ threads, the next round starts. If the current detection model is still busy in processing a previous frame, then the current frame will be dropped, and the detection results from the latest processed frame will be reused as the detection approximation for this dropped frame.
The video frame sequence synchronizer will enforce temporal input stream ordering over both the detection processed frames and the randomly dropped frames prior to displaying the live streaming of the detection processed output results via visualization or video streaming player. 
When the input video stream rate $\lambda$ is much faster than the per-model detection processing rate $\mu$, the round robin scheduler may incur larger amount of random frame dropping. A weighted round robin typically refers to a static resource adaptive scheduler. It assigns a higher weight to those detection models that run much faster, for example, due to high-capacity AI hardware known at edge device configuration, such as a high end GPU card instead of a low end Intel NCS2 stick. Compared to the round robin baseline, the static weighted RR scheduler can further improve the parallel detection throughput while maximizing execution parallelism.  

A {\bf first come and first serve (FCFS) scheduler} will assign incoming video frames to the $n$ detection models one at a time in the first come first serve order. Concretely, it starts by assigning $n$ input frames to the $n$ available detection models. Then it will assign the $(n+1)^{th}$ input frame to the first detection model that becomes available, instead of following a rigid round robin order for parallel detection processing. The FCFS scheduler is particularly suitable for the scenarios in which either the $n$ detection models or the $n$ AI hardware attachments are heterogeneous, and have different runtime performance for object detection processing. 

The {\bf performance aware proportional scheduler} is a dynamic runtime performance aware scheduler. It extends the round robin scheduler by periodically computing and assigning dynamic weights to the $n$ detection models. Concretely, upon assigning the first $n$ input frames to the $n$ detection models, it will start to monitor the execution time of each detection model and periodically compute the weight for each of the $n$ detection models based on the historical statistics on the execution performance of the $n$ models with respect to detection processing efficiency. Unlike the weighted round robin, which assigns a constant weight to each detection model at compile time, the performance-aware proportional scheduler dynamically assigns a weight to each detection model, enabling the parallel detection scheduling decision to take into account multiple dynamic factors that may impact on runtime performance, such as network bandwidth, edge device execution state, runtime state of AI hardware. 

Based on the chosen scheduling algorithm, the edge parallel model detection scheduler will schedule the $n$ threads, one for each detection model. The processed video frames will be sorted by the sequence synchronizer based on their original temporal streaming sequence for visualization and video player presentation of the detection processed frames, including detected objects in bounding boxes with class labels and confidence statistics. Note that for first come first serve and proportional scheduling algorithms, the parallel detection model scheduler will assign frames to threads in an opportunistic manner. The sequence synchronizer will be responsible for enforcing the temporal sequencing constraints on the detection processed frames before forwarding them to the visual presentation phase. Although with round-robin, the temporal sequence constraint is followed by the parallel detection scheduler, the sequence synchronizer serves as an additional checkpoint to ensure and enforce the temporal sequence of the output frames, which is important in the presence of random frame dropping.

\section{Experimental Analysis} \label{section:experimental-analysis}
We use two benchmark videos from the MOT-15 challenge dataset~\cite{MOTChallenge2015}: ADL-Rundle-6 and ETH-Sunnyday, as shown in Table~\ref{table:two-test-videos}. 
Both videos are taken in unconstrained real environments. ADL-Rundle-6 video is shot from a static camera at 30 FPS, has 525 frames in total, and the video resolution is 1920$\times$1080. ETH-Sunnyday is shot from a moving camera at 14 FPS with a resolution of 640$\times$480, and a total of 354 frames. We use two representative object detection models: SSD300~\cite{ssd} and YOLOv3~\cite{yolov3} for performing the object detection task on both videos. 

\newcommand{\smalltabfigure}[2]{\raisebox{-0.5\height}{\includegraphics[#1]{#2}}}
\begin{table}[h!]
\caption{{Two Test Videos}}
\label{table:two-test-videos}
\scalebox{0.95}{
\centering
\small
\begin{tabular}{|c|c|c|}
\hline
Video Name & ADL-Rundle-6 & ETH-Sunnyday \\ \hline
Video FPS & 30 & 14 \\ \hline
\#Frames & 525 & 354 \\ \hline
Resolution & 1920$\times$1080 & 640$\times$480 \\ \hline
Camera & static & moving \\ \hline
Example & \smalltabfigure{width=0.18\textwidth}{video-frames/ADL_000228}
&
\smalltabfigure{width=0.15\textwidth}{video-frames/ETH_000293}
\\ \hline
\end{tabular}
} 
\end{table}

\begin{table}[h!]
\caption{{Two Object Detection Models}}
\label{table:two-object-detection-models}
\scalebox{0.93}{
\centering
\small
\begin{tabular}{|c|c|c|c|c|}
\hline
Model & Backbone & Input Size & Model Size & Data Type \\ \hline
SSD300 & VGG-16 & 300$\times$300$\times$3 & 51MB & FP16 \\ \hline
YOLOv3 & DarkNet-53 & 416$\times$416$\times$3 & 119MB & FP16 \\ \hline
\end{tabular}
} 
\end{table}

To understand the impact of object detection models pre-trained using different DNN algorithms, we include SSD300 and YOLOv3 in our experiments, as shown in Table~\ref{table:two-object-detection-models}. 
Both SSD300 and YOLOv3 use their default backbone deep neural networks, i.e., VGG-16 for SSD300 and DarkNet-53 for YOLOv3. For detecting objects, the input video frame will be first resized to the input size of the object detection model, that is 300$\times$300$\times$3 for SSD300 and 416$\times$416$\times$3 for YOLOv3. Then, the deep neural network in SSD300 or YOLOv3 will perform inference on the resized video frame, followed by post-processing, such as non-maximum suppression (NMS), for extracting the bounding box, class label and object confidence for each detected object. SSD300 and YOLOv3 are converted into the FP16 (float16) data type to be deployed on Intel NCS2 sticks. 

Furthermore, to understand the impact of edge servers with different computational capacities, we include two types of edge servers as specified in Table~\ref{table:edge-server-configuration}. For high capacity type of edge servers, we configure the edge server with a fast CPU, Intel i7-10700K, with 24GB main memory and Ubuntu 20.04. For low capacity type of edge servers, we configure the edge server with an AMD A6-9225 CPU, with 12GB memory and Ubuntu 18.04. In both cases, a total of up to seven Intel NCS2 sticks are attached to these two edge servers via a USB 3.0 hub in all experiments reported in this paper.

\begin{table}[h!]
\centering
\caption{Edge Server Configuration}
\label{table:edge-server-configuration}
\scalebox{1.0}{
\centering
\begin{tabular}{|c|c|c|}
\hline
Edge Server & Fast & Slow \\ \hline
CPU & Intel i7-10700K & AMD A6-9225 \\ \hline
CPU Frequency & 3.8GHz & 2.6GHz \\ \hline
CPU \#Cores & 8 & 2 \\ \hline
Main Memory Size & 24GB & 12GB \\ \hline
OS & Ubuntu 20.04 & Ubuntu 18.04\\ \hline
\end{tabular}
} 
\end{table}

\begin{table*}[h!]
\centering
\caption{{Parallel Detection using Multiple NCS2 Sticks (ETH-Sunnyday)}}
\label{table:exp-multi-ncs2-stick-eth-sunnyday}
\scalebox{1.0}{
\centering
\begin{tabular}{|c|c|c|c|c|c|c|c|c|c|}
\hline
\multicolumn{2}{|c|}{Detection Processing} & Zero Frame Dropping & Single AI-Hardware & \multicolumn{6}{c|}{ Parallel Detection Models} \\ \hline
Model & \#NCS2 & 1 & 1 & 2 & 3 & 4 & 5 & 6 & 7 \\ \hline
\multirow{2}{*}{SSD300} & Detection FPS & 2.3 & 2.3 & 4.6 & 6.9 & 9.2 & 11.5 & 13.8 & 16.0 \\ \cline{2-10} 
 & mAP (\%) & 74.5 & 69.0 & \textbf{78.7} & \textbf{78.6} & \textbf{77.5} & \textbf{77.5} & \textbf{77.5} & \textbf{74.5} \\ \hline
\multirow{2}{*}{YOLOv3} & Detection FPS & 2.5 & 2.5 & 5.1 & 7.5 & 10.0 & 12.4 & 14.8 & 17.3 \\ \cline{2-10} 
 & mAP (\%) & 86.9 & 66.1 & 83.9 & 86.5 & 86.5 & 86.5 & \textbf{86.9} & \textbf{86.9} \\ \hline
\end{tabular}
} 
\end{table*}

\begin{table*}[h!]
\centering
\caption{{Parallel Detection using Multiple NCS2 Sticks (ADL-Rundle-6)}}
\label{table:exp-multi-ncs2-stick-adl-rundle-6}
\scalebox{1.0}{
\centering
\begin{tabular}{|c|c|c|c|c|c|c|c|c|c|}
\hline
\multicolumn{2}{|c|}{Detection Processing} & Zero Frame Dropping &  Single AI Hardware & \multicolumn{6}{c|}{Parallel} \\ \hline
Model & \#NCS2 & 1 & 1 & 2 & 3 & 4 & 5 & 6 & 7 \\ \hline
\multirow{2}{*}{SSD300} & Detection FPS & 2.3 & 2.3 & 4.6 & 6.9 & 9.1 & 11.5 & 13.7 & 16.0 \\ \cline{2-10} 
 & mAP (\%) & 54.4 & 46.7 & \textbf{56.2} & \textbf{55.8} & \textbf{55.4} & \textbf{55.7} & \textbf{55.7} & \textbf{54.7} \\ \hline
\multirow{2}{*}{YOLOv3} & Detection FPS & 2.5 & 2.5 & 5.1 & 7.5 & 10.0 & 12.5 & 14.8 & 17.3 \\ \cline{2-10} 
 & mAP (\%) & 62.5 & 42.7 & 56.7 & 61.2 & \textbf{62.7} & \textbf{62.7} & \textbf{62.7} & \textbf{62.7} \\ \hline
\end{tabular}
} 
\end{table*}

\subsection{Parallel Detection Effectiveness}

\noindent {\bf ETH-Sunnyday.} We first evaluate our multi-model detection parallel approach on ETH-Sunnyday by using up to seven Intel NCS2 sticks, and hence running up to seven detection models in parallel. Table~\ref{table:exp-multi-ncs2-stick-eth-sunnyday} shows the experimental results for two object detection models, SSD300 and YOLOv3. 
We include the zero frame dropping in the offline processing as the baseline for comparison with two scenarios: (1) the online object detection using only one detection model on one AI hardware (e.g., one NCS2 stick); and (2) the online parallel detection with $n$ models by varying $n$ from 2 to 7.

Three interesting observations are highlighted. (1) Our model parallel approach achieved almost linear scalability with the number of NCS2 sticks. Given one detection model runs on each of the 7 NCS2 sticks, the parallel detection with 7 models significantly improves the detection processing throughput ($\sigma_{P}$) from 2.3 FPS to 16.0 FPS for pre-trained SSD300 model and from 2.5 FPS to 17.3 FPS for pre-trained YOLOv3 detector. This represents 6.96$\times$ speed-up for SSD300 and 6.92$\times$ speed-up for YOLOv3, compared to using a single detection model on one NCS2 stick. 
(2) Based on our proposed method for determining the detection parallel parameter $n$ in Section~\ref{parameter_n}, for ETH-Sunnyday with the input video stream rate of 14 FPS, setting $n$ to be in the range of [$4, 6$] is sufficient for delivering near real time detection processing throughput. This experiment confirms this method. For $n$ NCS2 sticks, running $n$ YOLOv3 detection models in parallel, by varying $n$ from 4, 5 to 6, we can achieve a detection processing rate of 10.0 FPS, 12.4 FPS or 14.8 FPS respectively with mAP of 86.5\% or higher for all three cases. For street view of walking pedestrians, the moving camera capturing at 10 FPS or higher is considered comfortable and near real time for human perception. Hence, choosing $n$ by 4, 5, or 6 are considered acceptable configurations for our multi-model detection parallel approach.
(3) By choosing $n=7$, the parallel detection processing rate exceeds the input video stream rate of 14 FPS for both SSD300 and YOLOv3, achieving the same mAP: 74.5\% for SSD300 with 7 NCS2 sticks and 86.9\% for YOLOv3 as the zero frame dropping baseline. Clearly, adding one additional AI hardware will consume additional computation cost, including battery consumption if edge device is not wired, although the computation time is negligible thanks to parallel execution of all $n$ models.



\begin{figure}[h!]
\centering
    \includegraphics[width=0.45\textwidth]{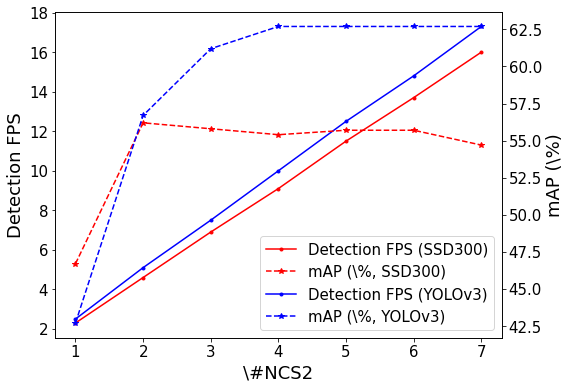}
    \caption{Detection FPS and mAP for Multiple NCS2 Sticks (ADL-Rundle-6, Video FPS=30)}
    \label{fig:ssd300-yolov3-multi-NCS2-adl-rundle-6}
\end{figure}

\noindent {\bf ADL-Rundle-6.} We next evaluate our parallel detection implementation on ADL-Rundle-6, another test video from MOT-15~\cite{MOTChallenge2015}, using the same experimental setup, including varying $n$ from 2 to 7 with up to 7 Intel NCS2 sticks and running SSD300 and YOLOv3 pre-trained models for object detection. Table~\ref{table:exp-multi-ncs2-stick-adl-rundle-6} 
shows the results. We highlight the following observations.
(1) Similar to ETH-Sunnyday in Table~\ref{table:exp-multi-ncs2-stick-eth-sunnyday}, the multi-model parallel detection approach achieves almost linear scalability over the number of NCS2 sticks on ADL-Rundle-6. (2) The input video stream rate for ADL-Rundle-6 is 30 FPS, doubling the rate of ETH-Sunnyday. However, comparing to the zero frame dropping baseline with mAP of 54.4\% and FPS of 2.3 for SSD pre-trained model and mAP of 62.5\% and FPS of 2.5 for YOLOv3 pre-trained model, when running parallel detection processing, choosing $n$ by 4, 5, 6 or 7 will offer better mAP and higher FPS: For SSD, when $n$ is set to 5, 6 or 7, we achieve the detection processing rate of 11.5, 13.7 and 16.0 respectively with mAP of 54.7\% or higher for these three settings. For YOLOv3, we achieve the detection processing rate of 10.0, 12.5, 14.8 and 17.3 respectively with mAP of 62.7\% for all four settings. (3) According to our method for determining the detection parallel parameter $n$, the choice of $n$ should be in the range of [$\lceil 10/2.3 \rceil, \lceil 30/2.3 \rceil$] for SSD, which is [$5, 14$], and in the range of $(\lceil 10/2.5 \rceil, \lceil 30/2.5 \rceil)$ for YOLOv3, which is [$4, 12$]. Even when we only have up to 7 Intel NCS2 sticks for this experiment, we show that with 4, 5, 6, or 7 detection models running in parallel, we can achieve near real time detection processing throughput higher than 10 FPS while maintaining mAP higher than the zero frame dropping baseline (measured using offline processing).

Figure~\ref{fig:ssd300-yolov3-multi-NCS2-adl-rundle-6} shows the trend of FPS (y-axis on the left) and mAP (y-axis on the right) for running SSD and YOLOv3 pre-trained models, by varying $n$ from 1 to 7 (x-axis). First, both SSD and YOLOv3 show the linear scalability with respect to the detection processing rate in FPS. Second, for YOLOv3, as $n$ increase from 1 to 4, the multi-model detection parallel approach continues to increase its mAP. As $n$ continues to increase from 4 to 7, the mAP is stabilized at 62.7\%. For SSD, the multi-model detection parallel approach always achieves mAP higher than the no-frame dropping baseline of 54.4\%, with $n\geq 2$. 
For SSD300 on ADL-Rundle-6, our multi-model parallel detection with 3 NCS2 sticks has the detection processing rate of 6.9 FPS. This results in a small amount of random frame dropping, which is on average $\lceil 30/6.9 - 1 \rceil = 4$ frames for each detection processed frame, compared to random dropping of 13 frames ($\lceil 30/2.3 - 1 \rceil = 13$) when using one detection model running on a single NCS2 stick.
For YOLOv3, when using 5 NCS2 sticks, the number of randomly dropped video frames decreases significantly from on average 11 frames for each detection processed frame ($\lceil 30/2.5 - 1 \rceil = 11$) when using one NCS2 stick to only dropping on average 2 frames ($\lceil 30/12.5 - 1 \rceil = 2$) for each detection processed frame. This also explains the reason for the improved mAP for both SSD300 and YOLOv3 since the number of dropped video frames is significantly reduced.


We provide a visualization comparison to show the effectiveness of our multi-model parallel detection approach with respect to the input video stream rates, accessible at \url{https://github.com/git-disl/EVA}.


\begin{table*}[h!]
\centering
\caption{{Comparison of Power Efficiency of Different Hardware Devices}}
\label{table:comparison-power-effieiency}
\scalebox{1.0}{
\begin{tabular}{|c|c|c|c|c|}
\hline
Power   Consumption & Intel   NCS2 & Slow   CPU (AMD   A6-9225) & Fast   CPU (Intel   i7-10700K) & GPU (GTX TITAN X) \\ \hline
TDP   (Watts) & 2~\cite{ncs2-power-consumption} & 15~\cite{amd-power-consumption} & 125~\cite{intel-power-consumption} & 250~\cite{nvidia-power-consumption} \\ \hline
Detection   FPS & 2.5 & 0.4 & 13.5 & 35~\cite{yolov3} \\ \hline
Detection   FPS / Watt & 1.25 & 0.03 & 0.11 & 0.14 \\ \hline
\end{tabular}
} 
\vspace{-2mm}
\end{table*}

\begin{table*}[h!]
\centering
\caption{{Experiments with RR and FCFS Scheduler (ETH-Sunnyday, YOLOv3)}}
\label{table:exp-rr-fcfs-eth-sunnyday-yolov3}
\scalebox{1.0}{
\begin{tabular}{|c|c|c|c|c|c|c|c|c|c|}
\hline
Detection   FPS & \#NCS2 & 0 (CPU Only) & 1 & 2 & 3 & 4 & 5 & 6 & 7 \\ \hline
\multirow{3}{*}{Round-Robin} & NCS2 Only & - & 2.5 & 5.1 & 7.5 & 10.0 & 12.4 & 14.8 & 17.3 \\ \cline{2-10} 
 & Fast CPU + NCS2 & 13.5 & 5.1 & 7.6 & 10.1 & 12.7 & 15.0 & 17.6 & 20.1 \\ \cline{2-10} 
 & Slow CPU + NCS2 & 0.4 & 0.9 & 1.3 & 1.8 & 2.2 & 2.6 & 3.1 & 3.4 \\ \hline
\multirow{3}{*}{FCFS} & NCS2 Only & - & 2.5 & 5.1 & 7.5 & 9.9 & 12.5 & 15.0 & 17.3 \\ \cline{2-10} 
 & Fast CPU +   NCS2 & 13.5 & 16.0 & 17.1 & 19.4 & 22.0 & 24.3 & 26.7 & 29.0 \\ \cline{2-10} 
 & Slow   CPU + NCS2 & 0.4 & 3.0 & 5.5 & 7.8 & 10.3 & 12.7 & 14.9 & 17.9 \\ \hline
\end{tabular}
}
\vspace{-2mm}
\end{table*}

\subsection{Energy Efficiency for Detection at Edge}

Edge devices are often with wireless connectivity and limited power supply. Therefore, energy efficiency is an important consideration for video analytics in edge systems. Table~\ref{table:comparison-power-effieiency} shows the power consumption of the AI hardware attached to the two types of edge servers (fast and slow) used in the experiments reported in this paper, including Nvidia GTX TITAN X (the popular GPU), Intel NCS2, AMD A6-9225 (the slow server), and Intel i7-10700K (the fast edge server). We run the YOLOv3 object detection model on these four types of edge devices and measure the energy consumption and FPS as reported in Table~\ref{table:comparison-power-effieiency}. We highlight two interesting observations.
{\it First}, different AI devices consume different amounts of power. We use the TDP (Thermal Design Power) to indicate the power dissipation of each AI device. For example, using one Intel NCS2 stick only consumes 2 watts. In comparison, using a single GTX TITAN X GPU has the highest energy consumption with 250 watts. When running YOLOv3 on edge server with slow CPU without AI hardware, it consumes 15 watts. In comparison, when running on edge server with fast CPU, it consumes 125 watts.   
{\it Second}, Intel NCS2 achieved the highest energy efficiency in terms of detection FPS per watt. Here we measure detection processing throughput in FPS with zero frame dropping for all four detection execution environments at edge. Although edge server equipped with Titan X GPU achieves 35 FPS, followed by the fast CPU with 13.5 FPS, Intel NCS2 with 2.5 FP. The slow CPU edge server has the slowest detection performance of 0.4 FPS. Based on both TDP (Watts) and FPS for all four settings, 
we next measure the energy efficiency in terms of detection FPS per watt, which is computed by dividing the detection FPS by the corresponding TDP. The edge device with Intel NCS2 offers the highest detection FPS per watt (1.25) compared to the edge device with GPU (0.14), edge device powered by fast CPU (0.11) and edge device powered by slow CPU (0.03).  
This further verifies that edge AI hardware such as Intel NCS2 stick is competitive in edge computing with high energy efficiency and suitable for deploying deep neural network models on wireless edge devices. 


\begin{table*}[h!]
\centering
\caption{{Comparison of Bandwidth for Different Interfaces}}
\label{table:bandwidth-comparison}
\scalebox{1.0}{
\begin{tabular}{|c|c|c|c|c|c|c|c|c|}
\hline
Port & USB 2.0 & USB 3.0 & Ethernet             & 10 Gigabit Ethernet & WiFi 6        & 4G (peak)        & 5G (peak)            \\ \hline
Bandwidth & 480 Mbps~\cite{usb-2-3-ethernet-speed} & 5 Gbps~\cite{usb-2-3-ethernet-speed}   & 100 Mbps$\sim$1 Gbps~\cite{usb-2-3-ethernet-speed} & 10 Gbps~\cite{role-10-gigabit-ethernet}  & $\sim$10 Gbps~\cite{wifi-6} & 1 Gbps ~\cite{cellnetwork-5g-4g} & 20 Gbps~\cite{cellnetwork-5g-4g} \\ \hline
\end{tabular}
} 
\end{table*}

\begin{table}[h!]
\centering
\caption{{\small The Impact of Connection Interface (ADL-Rundle-6)}}
\label{table:connection-interface}
\scalebox{0.962}{
\begin{tabular}{|c|c|c|c|c|c|c|c|c|}
\hline
\multicolumn{2}{|c|}{\#NCS2}      & 1   & 2   & 3   & 4    & 5    & 6    & 7    \\ \hline
\multirow{2}{*}{SSD300} & USB 2.0 & 2.0 & 3.9 & 5.9 & 7.8  & 9.7  & 11.6 & 13.2 \\ \cline{2-9} 
                        & USB 3.0 & 2.3 & 4.6 & 6.9 & 9.1  & 11.5 & 13.7 & 16.0 \\ \hline
\multirow{2}{*}{YOLOv3} & USB 2.0 & 1.9 & 3.7 & 5.5 & 7.2  & 8.1  & 8.0  & 8.1  \\ \cline{2-9} 
                        & USB 3.0 & 2.5 & 5.1 & 7.5 & 10.0 & 12.4 & 14.8 & 17.3 \\ \hline
\end{tabular}
} 
\end{table}

\subsection{Impact of Scheduling Algorithms}
This set of experiments is designed to measure the impact of using different parallel detection scheduling algorithms. We focus on comparing the baseline round robin (RR) scheduler with the first come first serve (FCFS) scheduler. Table~\ref{table:exp-rr-fcfs-eth-sunnyday-yolov3} reports the experimental results with these two schedulers by running YOLOv3 on the ETH-Sunnyday video. We highlight three interesting observations. 

{\it First}, consider NCS2 Only scenario, in which homogeneous AI hardware attachments are used with each running YOLOv3 on ETH-Sunnydy test video, the RR and FCFS schedulers show similar performance as expected when performing multi-model parallel detection using only $n$ NCS2 sticks by varying $n$ from 1 to 7, where the detection FPS for both RR and FCFS schedulers ranges from 2.5 to 17.3 FPS.

{\it Second}, when we carry out the multi-model parallel detection using heterogeneous edge devices, we expect different performance impact when using the RR schedule compared to using the FCFS scheduler. In the next experiment, we use Intel i7 (Fast CPU) as the edge server with $n$ NCS2 sticks by varying $n$ from 1 to 7, we perform the multi-model parallel detection using both fast CPU and the $n$ NCS2 sticks. We deploy YOLOv3 on CPU via the OpenVINO framework. This set of experiments shows that compared to the RR scheduler, the FCFS scheduling algorithm achieves much higher detection FPS with $n$ NCS2 sticks when varying $n$ from 1 to 7. This is because FCFS is in general more efficient for heterogeneous execution environments. Concretely, the fast CPU has much higher detection processing throughput (13.5 FPS) compared to that of a NCS2 stick (2.5 FPS). FCFS can effectively leverage such performance disparity whereas the RR scheduler fails to take into account such detection performance disparity when scheduling parallel detection execution. 

{\it Third}, with the Fast CPU as the edge server and 7 NCS2 sticks, the FCFS scheduler can consistently achieve higher throughput performance of 29 FPS, compared to the performance of 20.1 FPS by the RR scheduler under the same runtime setup. We next conduct the same set of experiments by replacing the fast CPU with slow CPU. It is observed that even with the slow CPU with poor detection of 0.4 FPS, using FCFS scheduler, our multi-model parallel detection approach can still deliver slightly better performance compared to using NCS2 only, because the slow CPU with NCS2 solution benefits from one additional model for parallel object detection. However, when using the RR scheduler, the slow CPU with NCS2 will suffer notably because the poor performance of the detection model running on the slow CPU had significantly negative impact on the overall performance of the parallel detection approach due to the large amount of random frame dropping at the slow CPU. 

In summary, FCFS scheduling algorithm provides more consistent performance for the multi-model parallel detection approach under both homogeneous and heterogeneous runtime environments. In our first prototype implementation, we set FCSC as the default scheduler.


\subsection{Impact of Connection Interface}

To install multiple AI hardware sticks to an edge server, one would need a USB hub. In our first prototype, we connect multiple Intel NCS2 sticks to the edge servers via the USB port with a USB hub.  Different versions of USB ports have different bandwidths. For instance, the bandwidth for USB 3.0 is 5 Gbps/s, which is much higher than USB 2.0 with about 0.5 Gbps/s~\cite{usb-2-3-ethernet-speed}. By connecting AI hardware such as an NCS2 stick to an edge server, the computation heavy DNN model for object detection will be hosted on the AI hardware for detection processing. Hence, the edge server will need to transfer the input video data to the attached NCS2 stick to perform the object detection processing. Hence, the connection bandwidth between the edge server and its attached AI hardware is critical. For example, we can connect the multiple Intel NCS2 sticks to the fast edge server via USB 2.0 or USB 3.0, and the two USB ports have different connection bandwidth, which subsequently impacts the detection processing throughput performance in FPS, as shown in  Table~\ref{table:connection-interface}. This set of experiments reports the results by running pre-trained object detection models by SSD300 and YOLOv3 on ADL-Rundle-6 test video. We highlight two interesting observations.

{\it First,} for both SSD300 and YOLOv3, edge detection with USB 3.0 achieves much higher performance in FPS than the same edge server with USB 2.0 port regardless whether we run object detection using only one NSC2 stick or using multiple NCS2 sticks. USB 3.0 port outperforms USB 2.0 port in both bandwidth and FPS throughput.
{\it Second}, the input data size for YOLOv3 is 416$\times$416$\times$3=519,168 which is about 2$\times$ of the input size of SSD300 with 300$\times$300$\times$3=270,000. YOLOv3 has higher bandwidth requirements than SSD300, which also explains that YOLOv3 fails to further improve the detection FPS after \#NCS2=5 with USB 2.0. 

This set of experiments motivates us to consider the high speed connection offered by 5G or 6G networks.  
Table~\ref{table:bandwidth-comparison} compares the network bandwidths with different types of connections. The reported peak bandwidth of 5G network could reach 20 Gbps~\cite{cellnetwork-5g-4g}, followed by 10 Gigabit Ethernet~\cite{role-10-gigabit-ethernet} and  WiFi 6 with 10 Gbps~\cite{wifi-6}. With guaranteed 10 Gigabit or higher network connectivity, our multi-model parallel detection approach can run effectively on a group of nearby edge nodes, with one edge node running one object detection model. However, with Ethernet~\cite{usb-2-3-ethernet-speed} or 4G network with 1 Gbps~\cite{cellnetwork-5g-4g} connection, using multiple AI hardware sticks on a single edge server with a USB 3.0 hub represents a more efficient approach in terms of both detection performance and energy consumption. 


\noindent {\bf Impact of Programming Languages on Detection Performance at Edge.\/}
During our prototype development for multi-model parallel object detection on edge devices, we also observed that when the parallel object detection model YOLOv3 is implemented in different programming languages, such as Python or C++, even with the same runtime environments, they achieve different detection throughput performance in FPS. Table~\ref{table:programming-language} reports the results from this set of experiments. 
Overall, the Python implementation fails to scale well with more than two NCS2 sticks, only achieving 9.7$\sim$9.8 FPS when $n$ varies from 3 to 7. In comparison, the C++ implementation scales as we increase the number of  NCS2 sticks from 2 to 7. The overall speedup of C++ implementation is 7$\times$ with 7 NCS2, showing some trend of linear scalability, though the synchronization overhead of the C++ implementation leads to slightly lower performance for 1 and 2 NCS2 sticks than the Python implementation. For the Python implementation, the current OpenVINO framework does not support multi-process in Python. For the Python multi-thread implementation, due to the Python global lock, the execution is ultimately serialized, consequently impairing the scalability.

\begin{table}[h!]
\centering
\caption{{\small The Impact of Programming Languages on parallel detection performance in FPS (YOLOv3, ADL-Rundle-6)}}
\label{table:programming-language}
\scalebox{1.0}{
\begin{tabular}{|c|c|c|c|c|c|c|c|}
\hline
\#NCS  & 1   & 2   & 3    & 4    & 5    & 6    & 7    \\ \hline
Python & 4.8 & 9.4 & 9.8  & 9.8  & 9.7  & 9.7  & 9.7  \\ \hline
C++    & 4.5 & 9.1 & 13.5 & 18.0 & 22.3 & 27.5 & 32.4 \\ \hline
\end{tabular}
} 
\end{table}


\section{Conclusion}

We have shown that even with well-trained DNN object detectors, the online detection quality in edge systems may deteriorate due to limited capacity for running DNN object detection models and random frame dropping when the frame processing rate is significantly slower than the incoming video stream rate. We present a multi-model parallel detection approach to effectively speed up the detection processing throughput at edge. We show that using the FCFS scheduler, our multi-model parallel detection approach can effectively speed up the detection throughput performance by minimizing the disparity in the detection processing rates on heterogeneous detection execution environments at edge. Extensive experiments are conducted using both SSD300 and YOLOv3 pre-trained DNN models on benchmark videos with different video stream rates. We show that our multi-model parallel detection approach can speed up the online object detection processing throughput (FPS) and deliver near real-time object detection performance at edge.


Our ongoing research includes implementing other performance aware model parallel scheduling algorithms, designing a multi-model and multi-device prototype system, capable of handling heterogeneous edge devices and heterogeneous detection models in terms of size and complexity. We are also interested in incorporating object tracking and segmentation techniques to take into consideration the temporal or spatial dependencies over individual video frames~\cite{object-tracking-survey,fast-online-object-tracking}.


\ifCLASSOPTIONcompsoc
  \section*{Acknowledgments}
\else
  \section*{Acknowledgment}
\fi

This work is partially sponsored by Cisco grant on edge computing. We thank the discussions and meetings with the Cisco team on edge computing. The authors from Georgia Institute of Technology are also partially supported by the National Science Foundation under Grants NSF 2038029 and NSF 1564097.



\bibliographystyle{IEEEtran}
\bibliography{IEEEabrv,reference}
%



\end{document}